\documentclass[aps,pre,groupedaddress,showpacs,preprint]{revtex4}
\usepackage{graphicx}
\usepackage[dvips]{epsfig}
\usepackage{dcolumn}
\usepackage{subfigure}%
\usepackage{bm}
\usepackage{amssymb}
\usepackage{amsmath}

\begin{document}

\title{Ergodic directional switching in mobile insect groups}

\author{Carlos Escudero$^1$, Christian A. Yates$^2$, Jerome Buhl$^3$, Iain D. Couzin$^4$, Radek Erban$^{2,5}$,
Ioannis G. Kevrekidis$^6$, and Philip K. Maini$^{2,7}$}

\affiliation{$1$ ICMAT (CSIC-UAM-UC3M-UCM), Departamento de
Matem\'{a}ticas, Facultad de Ciencias, Universidad Aut\'{o}noma de
Madrid, Ciudad Universitaria de Cantoblanco, 28049 Madrid, Spain \\
$2$ Centre for Mathematical Biology, Mathematical Institute,
University of Oxford, 24-29 St. Giles', Oxford, OX1 3LB, United Kingdom \\
$3$ School of Biological Sciences and Centre for Mathematical Biology, Heydon-Laurence Building, A08, The University of Sydney, NSW, Australia \\
$4$ Department of Ecology and Evolutionary Biology, Princeton
University, Princeton, NJ 08544, USA \\
$5$ Oxford Centre for Collaborative Applied Mathematics,
Mathematical Institute, University of Oxford, 24-29 St. Giles',
Oxford, OX1 3LB, United Kingdom \\
$6$ Department of Chemical Engineering, Program in Applied and
Computational Mathematics and Mathematics,
Princeton University, Princeton, NJ 08544, USA \\
$7$ Department of Biochemistry, Oxford Centre for Integrative
Systems Biology, University of Oxford, South Parks Road, Oxford
OX1 3QU, United Kingdom}

\begin{abstract}
We obtain a Fokker-Planck equation describing experimental data on
the collective motion of locusts. The noise is of internal origin
and due to the discrete character and finite number of
constituents of the swarm. The stationary probability distribution
shows a rich phenomenology including non-monotonic behavior of
several order/disorder transition indicators in noise intensity.
This complex behavior arises naturally as a result of the
randomness in the system. Its counterintuitive character
challenges standard interpretations of noise induced transitions
and calls for an extension of this theory in order to capture the
behavior of certain classes of biologically motivated models. Our
results suggest that the collective switches of the group's
direction of motion might be due to a random ergodic effect and,
as such, they are inherent to group formation.
\end{abstract}


\maketitle

\section{Introduction}

Emergence can be defined as the appearance of rich structures on a
large scale resulting from a multiplicity of simple interactions
at a considerably smaller scale. Collective animal motion is a
paradigmatic example of such an emergent phenomenon. Depending on
the species, there may exist different hierarchical levels that
determine how collective displacements are realized. For example,
in primate groups, an individual's dominance status can affect its
role in initializing collective movement. In the case of swarming
locusts no such hierarchies are present; the ability of each
individual to guide the band appears to be distributed relatively
evenly throughout the insect group. Herein we will concentrate on
groups of wingless locust nymphs which form marching bands rather
than flying swarms \cite{buhl}. The onset of collective motion in
locusts was experimentally demonstrated in \cite{buhl}, where it
was shown that sufficiently large insect densities placed in a
ring-shaped arena gave rise to a coherent displacement of the
band. Low densities were characterized by random dispersal of the
individuals, while for intermediate densities the coherent motion
was interrupted by sudden changes of direction (hereafter referred
to as ``switches''). This phenomenology was partially rationalized
by means of an adapted model based on that of Czir\'{o}k {\it et
al.} \cite{vicsek}, who formulate a paradigmatic model for
collective animal behavior in one dimension. In their original
model the position, $x_i$, and velocity, $u_i$, of locust $i$ are
evolved using the following two rules, identical for each
individual, $i=1,\dots, N$,
\begin{align}
x_i(t+1)&=x_i(t)+v_0u_i(t)\nonumber,\\
u_i(t+1)&=G(\bar{u}_i)+\xi_i\nonumber,
\end{align}
where $N$  is the total number of locusts. Here $\bar{u}_i$ is the
mean of the nondimensionalised velocities of locusts within a
certain radius, $R$, of the position, $x_i$, of locust $i$. The
function $G$ is such that $G(u)=(1+K)^{-1}[u+K \,
\mathrm{sgn}(u)]$ for a positive constant $K$, where
$\mathrm{sgn}(u)$ denotes the sign of $u$. The role of $G$ is to
adjust the average nondimensionalised velocity perceived by each
particle towards unity. $v_0$ is a constant associated with the
chosen time scale and $\xi_i$ is a random number drawn from the
uniform distribution in $[-\eta/2,\eta/2]$. The adapted version of
the model used in \cite{buhl} to model the movements of locust
nymphs in a quasi-one-dimensional arena takes the form
\begin{equation}
\label{uczirok} \frac{\mbox{d}x_i}{\mbox{d}t} = u_i, \qquad
\mbox{d}u_i = [G(\bar{u}_i)-u_i]\mbox{d}t + \beta_1 \mbox{d}W_i,
\qquad \text{for  } i=1, \dots, N,
\end{equation}
where $\mbox{d}W_i$ denotes the increments of independent Wiener
processes, $\beta_1$ is a positive constant describing the
amplitude of the noise and the function $G$ is as above.

A biologically motivated refinement of the model described by
Eq.~\eqref{uczirok} was given in \cite{yates}, where it was
postulated that individual locusts increase the randomness of
their movements in response to a loss of group alignment. This
behavior is the result of a particular multiplicative form of the
noise term (see Eq. \eqref{fp}), as opposed to the additive noise
in Eq. \eqref{uczirok}; this characteristic was shown to increase
the coherence of the group motion and to reduce the frequency of
direction switches \cite{yates}. The key point in the analysis
performed in \cite{yates} was the estimation of coefficients of an
effective Fokker-Planck equation (FPE) \cite{kevrekidis}, which is
written in terms of a macroscopic (low-dimensional) observable
\cite{varfree}, the average velocity of the marching group,
derived directly from the experimental data. In the present work
we approximate the drift and diffusion coefficients of the
effective FPE by analytical functions. This permits a more
thorough analysis and fosters further understanding of collective
dynamics of locusts. In addition we compare our results with those
of Eq.~\eqref{uczirok}, and discuss the disparities between the
two models.

\section{The model}

Coarse-grained analysis \cite{kevrekidis} allows us to obtain an
effective FPE describing the collective behavior of the locusts at
the macroscopic level. By using this coarse-graining technique
(see \cite{yates}) we were able to extract the coefficients of the
assumed underlying FPE describing the alignment of the locusts
from the experimental data presented in [1]. This approach enables
us to reduce our system - comprising a large number of degrees of
freedom - to a single collective variable, $u$, (referred to
variably, hereafter, as `alignment' or `average velocity') which
characterizes the system's macroscopic behavior. The proposed FPE
has a simple form and it can be expressed as
\begin{equation} \label{fp}
\partial_t P = -\alpha_2 \partial_u \left[ \left( u-\frac{u^3}{1-u^2} \right) P
\right]+ {\frac{\beta_2}{N}} \partial_{uu}[(1-u^2) P],
\end{equation}
for the probability $P(u,t)\mbox{d}u \mbox{d}t$ of finding the
system with an average velocity in the interval $(u,u+ \mbox{d}u)$
during the time interval $(t,t+\mbox{d}t)$; note that the
experimental situation in \cite{buhl} is quasi-one-dimensional,
allowing the use of a one-dimensional FPE \cite{yates}. We note
that this FPE corresponds to the following Langevin equation for
the average velocity
\begin{equation}
\label{langevin} \mbox{d}u = \alpha_2 \left( u-\frac{u^3}{1-u^2}
\right) \mbox{d}t + \sqrt{\frac{2 \beta_2}{N}} \sqrt{1-u^2} \,
\mbox{d}W,
\end{equation}
where $\mbox{d}W$ denotes the increments of a Wiener process and
the multiplicative noise is interpreted in the It\^{o} sense, as
prescribed by the experimentally obtained FPE~(\ref{fp}). In these
equations the average velocity $u$ is dimensionless and takes its
values in the interval $[-1,1]$. The values such that $|u|=1$
characterize the ideal situation in which all locusts march in
perfect coherence; the sign determines the direction. Of course,
formally substituting $|u|=1$ in Eq.~(\ref{langevin}) produces a
divergence in the drift, so for practical reasons one has to
assume that coherent motion implies $|u| \lesssim 1$ rather than a
strict equality. The value $u=0$ characterizes a total disorder;
realistic values of the average velocity lie between these two
extreme cases. Although the average velocity is a dimensionless
quantity in Eqs.~(\ref{fp}) and (\ref{langevin}), time is not.
Consequently $\alpha_2$ and $\beta_2$ have the dimensions of
time$^{-1}$. We shall estimate in the following their numerical
values using the experimental data from~\cite{buhl} and express
them in units of seconds$^{-1}$.

The proposed FPE \eqref{fp} describing the alignment
is a reasonably accurate approximation to the unknown FPE assumed
to underly the motion of the locusts, which captures their
experimental swarming behavior. It should be noted that such an
equation can only be obtained if the system being studied is
amenable to this sort of reduction.

For asymptotically large values of $N$, parameter $\alpha_2^{-1}$
(in equation ~\eqref{fp}) denotes the order of magnitude of the
relaxation time characterizing how long it takes the entire group
to become ordered when starting from a disordered configuration,
and $N\beta_2^{-1}$ indicates the order of magnitude of the
characteristic time over which the fluctuations of the mean
velocity develop. For the range of experimentally considered
locust numbers ($5\leq N \leq 40$) the observed values of
$\alpha_2$ and $\beta_2$ are approximately constant while we
expect the presence of a boundary layer for smaller values of $N$.
Since our results in \cite{yates} are rather noisy our goal is to
fit the order of magnitude of the model parameters instead of
attempting to obtain precise estimates. Comparing the proposed
analytical coefficients of Eq. \eqref{fp} to those obtained in
\cite{yates}, from the experimental data in \cite{buhl}, we obtain
$\beta_2/\alpha_2=2.4 \pm 1.7$. Employing the mean switching time
measurements in \cite{yates} we find $\alpha_2 = (6.65 \pm 2.63)
10^{-4} s^{-1}$ and $\beta_2 = (1.62 \pm 0.52) 10^{-3} s^{-1}$.

The FPE corresponding to Eq.~(\ref{uczirok}) can be obtained as a mean-field approximation,
\begin{equation} \label{vicsek} \partial_t P = - \alpha_1 \partial_u
\{ [\mathrm{sgn}(u)-u] P \} + \frac{\beta_1}{N} \partial_{uu} P,
\end{equation} where $\alpha_1 = K/(1+K)$ and $K$ is defined as for the function $G$ in Eq.~(\ref{uczirok}). The stationary solution of the FPE
~(\ref{vicsek}) can be derived as follows:
\begin{equation}
P_s(u)= \frac{\sqrt{\frac{\alpha_1 N}{2\pi \beta_1}}\exp \left(
-\frac{\alpha_1 N}{2 \beta_1} \right)}{1+\mathrm{erf} \left( \sqrt{
\frac{\alpha_1 N}{2 \beta_1}} \right)} \exp \left[ \frac{\alpha_1
N}{\beta_1} \left( |u|-{\frac{1}{2}}u^2 \right) \right]\label{equation:revisedmodelspd}.
\end{equation}
The values of the two maxima of this stationary probability
distribution (SPD) $u_{\max}=\pm 1$ and the minimum $u_{\min}=0$
are independent of the parameter values. This type of system has
been considered many times in the literature~\cite{hanggi,jung},
and we include it here simply for completeness and comparison with
the refined model~(\ref{langevin}).

In the absence of sources and sinks of probability, we can also
derive the SPD of the experimentally motivated FPE~\eqref{fp}:
\begin{equation}
\label{sprobability} P_s(u)= \mathcal{N} (1-u^2)^{-1-N
\alpha_2/\beta_2} \exp \left[ -\frac{N \alpha_2/(2 \beta_2)}{1-u^2}
\right],
\end{equation}
where $\mathcal{N}^{-1}= \int_{-1}^1 (1-u^2)^{-1-N
\alpha_2/\beta_2} \exp \left[ -N \alpha_2/[2 \beta_2(1-u^2)]
\right] \mbox{d}u$ is the inverse of the normalization constant.
This SPD is bounded, compactly supported in $[-1,1]$ and bimodal
for all values of the parameters.

Noise induced transitions have been studied traditionally by means
of the dynamics of the extrema of the SPD \cite{lefever}. For the
biologically motivated FPE \eqref{fp} the SPD shows one minimum
always located at $u_{\min}=0$, and two maxima at $u_{\max} = \pm
\sqrt{\alpha_2 + 2\beta_2/N}/\sqrt{2\alpha_2+2\beta_2/N}$, which
exist for all parameter values. One immediately notes $|u_{\max}|
\in (1/\sqrt{2},1)$, a fact related to the shape of the
``deterministic potential'' (the potential in the absence of
noise), which is the negative integral of the drift coefficient,
\begin{equation}
\mathcal{V}(u)=-\int_0^u \alpha_2 \left( s-\frac{s^3}{1-s^2} \right)
\,\mbox{d}s= -\alpha_2 \left[ u^2+{\frac{1}{2}} \ln(1-u^2) \right].
\end{equation}
This potential is bistable with one maximum located at the origin
and two minima at $\pm 1/\sqrt{2}$ independent of the parameter
values. For increasing noise intensity the probability maxima of
the SPD~\eqref{sprobability} (corresponding to the biologically
motivated FPE \eqref{fp}) separate from the deterministic
potential minima $\pm 1/\sqrt{2}$ and approach the boundary points
$\pm 1$. These facets of the SPD, when considered in the context
of the classical theory of noise induced transitions, imply that
the system is becoming ordered \cite{lefever}: the SPD maxima,
representing the states in which the system will most likely be
found, are further apart and thus there is a clearer
differentiation among those states. However, the experimental
evidence, based on switching times which decrease as the noise
magnitude increases, reveals that the system becomes disordered
\cite{buhl}. This indicates that for complex systems, restricting
the characterization of the dynamics to observations of the
evolution of the extrema may not be adequate in some
experimentally motivated situations. Herein we will try to carry
out a more complete characterization.

\section{Barrier Height}

Another indicator of order/disorder is the barrier height of the
effective potential. For the model given by Eq.~(1) the barrier
height decreases monotonically as the noise intensity increases as
can be seen from
Eq.~(\ref{equations:originalmodelpotentialandbarrier})
\begin{equation} \mathcal{V}_{\mathrm{eff}}(u)
= - \frac{\alpha_1 N}{\beta_1} \left( |u|-{\frac{1}{2}}u^2
\right), \qquad \Delta \mathcal{V}_{\mathrm{eff}} \equiv
\mathcal{V}_{\mathrm{eff}}(u_{\min})-\mathcal{V}_{\mathrm{eff}}(u_{\max})
= \frac{\alpha_1 N}{2
\beta_1},\label{equations:originalmodelpotentialandbarrier}
\end{equation}
where $\mathcal{V}_{\mathrm{eff}}$ is the effective potential and
$\Delta \mathcal{V}_{\mathrm{eff}}$ the corresponding barrier
height.

\begin{figure}
\centering
\subfigure[]{
\includegraphics[width=0.4\textwidth]{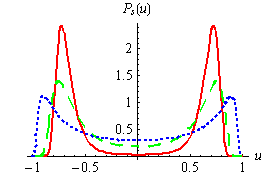}
\label{figure:SPDrevisedmodel}}
\subfigure[]{
\includegraphics[width=0.4\textwidth]{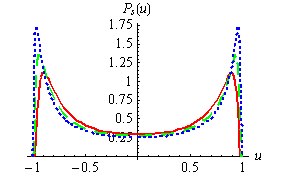}
\label{figure:SPDczirokmodel}}
\subfigure[]{
\includegraphics[width=0.4\textwidth]{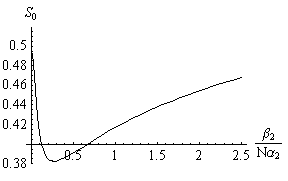}
\label{figure:secondmoment0}}
\subfigure[]{
\includegraphics[width=0.4\textwidth]{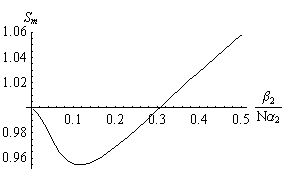}
\label{figure:secondmomentmax} } \caption{(Color online) Panels
\subref{figure:SPDrevisedmodel} and \subref{figure:SPDczirokmodel}
show the profile of the SPD, $P_s(u)$, from Eq.
\eqref{sprobability} plotted against the normalized mean velocity
$u$ for a varying noise intensity $\beta_2/(N\alpha_2)$. In panel
\subref{figure:SPDrevisedmodel} the solid red line represents
$\beta_2/(N\alpha_2)=0.05$, the dashed green line represents
$\beta_2/(N\alpha_2)=0.15$ and the dotted blue line represents
$\beta_2/(N\alpha_2)=1.5$. In panel \subref{figure:SPDczirokmodel}
the solid red line represents $\beta_2/(N\alpha_2)=1.5$, the
dashed green line represents $\beta_2/(N\alpha_2)=3.0$ and the
dotted blue line represents $\beta_2/(N\alpha_2)=5.5$. Panel
\subref{figure:secondmoment0} displays the second moment of the
revised model, centered at
 the origin, $S_0$, versus noise
strength $\beta_2/(N\alpha_2)$ and panel
\subref{figure:secondmomentmax} shows the second moment at a maximum, $S_m$, versus noise
strength $\beta_2/(N\alpha_2)$. The minima are attained for
$\beta_2/(N\alpha_2) \approx 0.27$ ($S_0$) and for
$\beta_2/(N\alpha_2)\approx 0.12$ ($S_m$).} \label{spdf}
\end{figure}

The effective potential for our revised model  (Eq.
\eqref{fp}) is given as
\begin{equation} \mathcal{V}_{\mathrm{eff}}(u) \equiv
-\ln [P_s(u)] = \frac{N\alpha_2}{2\beta_2(1-u^2)}+\left(
1+\frac{N\alpha_2}{2\beta_2} \right)\ln(1-u^2),
\end{equation}
 and the corresponding barrier height is
\begin{equation}
\Delta \mathcal{V}_{\mathrm{eff}} \equiv
\mathcal{V}_{\mathrm{eff}}(u_{\min})-\mathcal{V}_{\mathrm{eff}}(u_{\max})
= -1-\frac{N\alpha_2}{2\beta_2}+\left( 1+\frac{N\alpha_2}{\beta_2} \right)
\ln \left(2+ \frac{2\beta_2}{N\alpha_2} \right).
\end{equation}
As a function of noise intensity the barrier height exhibits a
minimum at $\left[\beta_2/(N\alpha_2)\right]_{\min} \approx 0.76$.
This means that for
$\beta_2/(N\alpha_2)<[\beta_2/(N\alpha_2)]_{\min}$ (sub-threshold)
the barrier height diminishes for stronger noise, but for
$\beta_2/(N\alpha_2)>[\beta_2/(N\alpha_2)]_{\min}$
(super-threshold) it increases as the noise strength grows.
Indeed, $\Delta \mathcal{V}_{\mathrm{eff}} \approx
[\ln(2)-1/2](N\alpha_2/\beta_2)$ when $N\alpha_2/\beta_2 \to
\infty$ and $\Delta \mathcal{V}_{\mathrm{eff}} \approx
-\ln(N\alpha_2/\beta_2)$ when $N\alpha_2/\beta_2 \to 0$. This
suggests that, while increased noise causes the system to become
more disordered for sub-threshold noise intensities,
super-threshold intensities might cause the system to become more
ordered as the noise grows. In short, the `barrier height' order
parameter shows a clear non-monotonicity when considered as a
function of noise strength. This appears like a counterintuitive
reentrant behavior, where the noise can have an ordering effect
for supercritical intensities~\cite{lefever}. Although this
behavior is interesting in itself, it is not biologically
relevant, as it requires a number of individuals $N \approx 3$,
beyond the validity of the model. Both characteristics of the SPD
(\ref{sprobability}) of our refined model, displacement of the
location of the maxima and non-monotonic variation of the barrier
height, can be seen in Fig. \ref{spdf}
\subref{figure:SPDrevisedmodel}-\subref{figure:SPDczirokmodel}.

Let us note that noise induced non-equilibrium phase
transitions~\cite{broeck,kawai} as well as other noise mediated
ordering phenomena~\cite{sancho} have been exhaustively studied in
the literature. This includes the detailed study of reentrant
transitions ~\cite{sailer}. However, most of these approaches have
assumed multiplicative noise interpreted in the Stratonovich
sense. The Stratonovich interpretation is able to destabilize a
deterministically stable state and in this way produce phase
transitions or other noise induced phenomena. The key technical
point in these cases is the appearance of a systematic
contribution to the deterministic dynamics coming from the noise
term, the so-called ``Stratonovich drift''~\cite{lefever}.
Therefore noise interpretation plays a fundamental role in the
development of these types of phenomena: in particular, many noise
induced phenomena are not possible if the noise interpretation is
that of It\^{o}. In this respect, our results are fundamentally
different as we only consider the It\^{o} interpretation for our
Langevin equation with multiplicative noise~(\ref{langevin}). We
also note that noise induced phase transitions which are
independent of the noise interpretation have also been
studied~\cite{carrillo}, but in much less detail. The mechanisms
leading to these phase transitions are based on the bifurcation of
the minima of an effective potential due to noise and, as such,
constitute the natural extension of noise induced
transitions~\cite{lefever} to spatially extended systems. Our
results relate to a zero dimensional system as do those
in~\cite{lefever}, but they are significantly different as the
model defined by Eq.~(\ref{langevin}) does not describe this kind
of bifurcation.

\section{Mean Switching Time}

We can further explore the model properties by considering the
mean switching time, $T(u)$, defined as the first time, on
average, that the alignment of the system, $u$, initialized such
that $-1<u<0$, reaches the origin ($u=0$). For our revised model
the moments of the switching time distribution are given,
recursively, by the solution of the equation
\begin{equation}
\label{mfptn} \alpha_2 \left( u-\frac{u^3}{1-u^2} \right)
\partial_u T_n + {\frac{\beta_2}{N}} (1-u^2) \partial_{uu} T_n=-nT_{n-1},
\end{equation}
subject to the boundary conditions $T(0)=0$ and $T'(-1)=0$, where
$T_n$ is the $n^{th}$ moment, correspondingly, $T \equiv T_1$ is
the mean switching time and $T_0 \equiv 1$. The second boundary
condition represents zero probability flux through $u=-1$. Eq.
\eqref{mfptn} is directly derived from the FPE using methods from
\cite{gardiner}. The solution to this equation for $n=1$ is
\begin{equation}
T(u)=\frac{N}{\beta_2}\int_u^0 \exp \left[
\frac{N\alpha_2/(2\beta_2)}{1-v^2}
\right](1-v^2)^{N\alpha_2/\beta_2}\int_{-1}^v \exp \left[
-\frac{N\alpha_2/(2\beta_2)}{1-w^2}
\right](1-w^2)^{-1-N\alpha_2/\beta_2} \mbox{d}w \mbox{d}v.
\nonumber
\end{equation}
This expression appears complicated, but one can derive its
asymptotic expansion for large values of $N\alpha_2/\beta_2$
(which implies large $N$ as $\alpha_2$ and $\beta_2$ are
approximately constant) by means of a Kramers-like
approximation~\cite{escudero}. It has the simple form
$T\left(-1/\sqrt{2}\right) \approx {\frac{\sqrt{2}\pi}{
\alpha_2}}\left(\frac{2}{\sqrt{e}}\right)^{N\alpha_2/\beta_2}$,
which reveals a pure exponential growth in the inverse noise
intensity $N\alpha_2/\beta_2$ for asymptotically large values.
Further moments of the switching time distribution can be
calculated from Eq.~\eqref{mfptn} for $n>1$. In the limit $N \to
\infty$ one finds the relation $T_n=n!T^n$. This relationship
implies, in turn, that the switching process is a Poisson process.

We can also compute the first passage time for the model
Eq.~\eqref{uczirok}. In this case we solve the equation $\alpha_1
[\mathrm{sgn}(u)-u]\partial_u T +
\frac{\beta_1}{N}\partial_{uu}T=-1$, subject to the boundary
conditions $T(0)=0$ and $T'(-\infty)=0$, where the latter
condition is the analogue of the previous zero flux condition
adapted for an SPD with infinite support. We find
\begin{equation}
\nonumber T(u)=\sqrt{\frac{\pi N}{2 \alpha_1 \beta_1}} \int_u^0
\exp \left[ \frac{N \alpha_1}{2 \beta_1}(1+v)^2 \right] \left\{
2-\mathrm{erfc} \left[ \sqrt{\frac{\alpha_1 N}{2 \beta_1}}(1+v)
\right] \right\} \mbox{d}v,
\end{equation}
which also behaves exponentially in $N$ for large values of $N$
but this time with an $N$ dependent prefactor (see Supplementary
Information of \cite{yates}). The relation between these two mean
switching times (the model Eq.~\eqref{uczirok} and that of the
revised model \cite{yates}) is extensively discussed in
\cite{yates}, so we will not reproduce the discussion here.

Now we compare the theoretical results with the exponential
fitting we have performed on experimental data from \cite{buhl}
for both the first and second moments of the switching time
distribution. The data are insufficient for us to reliably obtain
any moments higher than the second. For the mean switching time
$T$ and second moment $T_2$ we found
\begin{equation}
\nonumber
T = (970 \pm 120) \exp[(0.045 \pm 0.007)N] \, s, \quad
\sqrt{T_2/2} = (1300 \pm 190) \exp[(0.041 \pm 0.008)N] \, s.
\end{equation}
According to the relation $T_n=n!T^n$ for the moments of the
exponential distribution, these two values should be the same if
the switching process were Poissonian. Note that the exponential
growth is the same for both (within errors), while the prefactor
is larger for the second moment. This suggests that the switching
process is Poissonian for large $N$, that is, the probability
distribution for the switching events is
$\mathcal{P}=T^{-1}\exp(-t/T)$. For small values of $N$ the
behavior is more stochastic, as signaled by the larger prefactor
of the second moment (when $N$ is small the prefactor dominates
over the exponential). If the switching process is Poissonian then
this has a series of consequences concerning predictability: the
standard deviation being equal to the mean implies a 100\% error
in predictions. Furthermore, switching events are uncorrelated and
the distribution tail falls off exponentially for long times. This
allows for a higher probability of rare events than would be
allowed by a Gaussian tail. This also implies that the switching
process is Markovian, as predicted by the FPE. This can be seen
from the double-welled FPE (\ref{fp}) in the large $N$ limit.
After a short time the system relaxes to one potential minimum
where it stays an exponentially long time until the switch occurs.
Since practically all switches start at the minimum this erases
the memory and the Markov property is recovered. The verification
of this theoretical prediction by the experimental data suggests
that no important correlations have been suppressed in the
coarse-grained computation in~\cite{yates}, and that this method,
and the FPE (\ref{fp}), are suitable to describe the locust
dynamics exhibited by the experimental data.

\section{Second Moments}

Another indicator of the stochastic properties of the system is
the second moment, which measures the spread of the mean velocity,
$u$, with respect to some reference value. We consider two
variants, one centered at the origin $S_0 \equiv \int_{-1}^1 u^2
P_s(u) \mbox{d}u$, and one centered at one of the maxima of the
probability distribution $S_m \equiv \int_{-1}^1 (u-u_{\max})^2
P_s(u) \mbox{d}u$. Of course, the value of $S_m$ is the same for
both maxima as a consequence of the symmetry of the system. These
integrals have been computed numerically and are represented in
Fig. \ref{spdf}\subref{figure:secondmoment0}, centered at the
origin, and \ref{spdf}\subref{figure:secondmomentmax} centered at
a maximum. Both show non-monotonic behavior in noise intensity,
but attain their minima for different values of the noise
amplitude. This non-monotonic behavior, as well as the behavior of
the effective barrier height, are not reflected in the
relationship between mean switching time and the size of the noise
parameter ($N\alpha_2/\beta_2$): the mean switching time grows
monotonically with noise amplitude. For comparison we note that
both moments $S_0$ and $S_m$ grow monotonically with the inverse
noise intensity in the model given by Eq.~\eqref{uczirok}; in this
case they are
\begin{equation}
S_0 = 1+\frac{\beta_1}{\alpha_1 N}+\frac{\sqrt{\frac{2\beta_1}{\pi
\alpha_1 N}}\exp \left(-\frac{\alpha_1 N}{2 \beta_1} \right)}
{1+\mathrm{erf}\left( \sqrt{\frac{\alpha_1 N}{2 \beta_1}} \right)},
\qquad S_m = 2+\frac{\beta_1}{\alpha_1
N}+\frac{\sqrt{\frac{2\beta_1}{\pi \alpha_1 N}}\exp
\left(-\frac{\alpha_1 N}{2 \beta_1} \right)} {1+\mathrm{erf}\left(
\sqrt{\frac{\alpha_1 N}{2 \beta_1}} \right)}.
\end{equation}
There is another feature of the second moments of the revised
model \cite{yates}, in addition to the non-monotonic behavior,
that reveals new characteristics of the collective motion of
locusts not reflected by the model Eq.~\eqref{uczirok}. In this
model a reduction in the number of individuals increases the
values of both second moments. In the stronger noise situation the
probability distribution tails grow, which implies that there are
more individuals with a higher (absolute value) velocity. In our
case the probability is compactly supported in $[-1,1]$, as a
consequence of the biological fact that the propagation cannot be
better than perfect. For realistic values of the parameters the
system is in the weak noise regime (see Fig.
\ref{spdf}\subref{figure:secondmoment0}). This means that the
second moment centered at the origin decreases for a decreasing
number of locusts, exactly the opposite trend to that of the model
Eq.~\eqref{uczirok}. The reason is that the probability of finding
the system in the neighborhood of $u=0$ grows considerably for
stronger noise (as reflected by the decreasing barrier height),
largely compensating for the drift of the maxima towards the
boundaries of the support of the SPD. The experimentally derived
value of $\beta_2/N\alpha_2 = 0.12 \pm 0.08$ for $N=20$ agrees
with the minimizing value of the second moment centered at a
maximum, $\beta_2/N\alpha_2 \approx 0.12$ (see Fig.
\ref{spdf}\subref{figure:secondmomentmax}). This implies that its
behavior is not very sensitive to small changes in the number of
locusts.

\section{Conclusions}

We have seen that the FPE obtained from the coarse-grained
analysis of experimental data on the movement of locusts shows an
interesting phenomenology. Different indicators of order/disorder
may vary non-monotonically with noise intensity, possibly in a
contradictory manner. These findings reveal that these indicators
might not be suitable for the biologically motivated models
studied in this paper. We have also shown that the direction
switches are independently distributed for large numbers of
individuals. This makes them almost unpredictable from a practical
viewpoint. It seems that directional switches are produced by an
accumulation of errors (made by the locusts when trying to adapt
their velocity to that of their neighbors) that ordinarily
interfere and cancel each other out but, over exponentially long
times, have the possibility of accumulating and producing a
switch. According to the results presented here, specifically the
confirmation of the Poissonian character of the switching events,
it seems possible that directional switches are produced as a
consequence of the ergodic random evolution of the system. We note
the similarity of this process with Ising model ergodic
magnetization changes~\cite{brendel}. Indeed, the model of
Eq.~(\ref{uczirok}) can be thought of as an Ising model with
moving spins. It seems that the ergodic nature of the finite size
Ising model is preserved despite introducing movement of the
spins. More importantly it seems that this is a plausible
explanation, in the absence of external stimuli, of the sudden
changes of direction observed in animal groups.

\section*{Acknowledgements}

The authors are grateful to David Sumpter for useful comments and
discussions. This work was supported by the Oxford-Princeton
Research Partnership grant. CE acknowledges support by the MICINN
(Spain) through Project No. MTM2008-03754. CAY thanks EPSRC for
funding via the Systems Biology Doctoral Training Centre,
University of Oxford. JB was funded by the Australian Research
Council (ARC) Linkage and Discovery programs. IDC acknowledges
support from the Searle Scholars Program (08-SPP-201), National
Science Foundation (PHY-0848755), Office of Naval Research
(N00014-09-1-1074) and a DARPA Grant (HR0011-05-1-0057). This
publication was based on work (RE) supported in part by Award No
KUK-C1-013-04, made by King Abdullah University of Science and
Technology (KAUST); RE also thanks Somerville College, Oxford for
a Fulford Junior Research Fellowship. The research leading to
these results has received funding from the European Research
Council under the {\it European Community's} Seventh Framework
Programme ({\it FP7/2007-2013})/ ERC {\it grant agreement} No.
239870. IGK was partially supported by the AFOSR. PKM was
partially supported by a Royal Society Wolfson Research Merit
Award.

\end{document}